\documentclass[prl,twocolumn,showpacs,aps]{revtex4}
\usepackage{amsfonts,amssymb,amsmath,bm}
\usepackage[dvips]{graphicx}
\usepackage[dvips]{color}
\usepackage{times}

\begin{document}

\title{Dynamics of nearly spherical vesicles in an external flow}

\author{V. V. Lebedev, K. S. Turitsyn, and S. S. Vergeles}

\affiliation{Landau Institute for Theoretical Physics,
 Moscow, Kosygina 2, 119334, Russia}

\date{\today}

 \begin{abstract}

We analytically derive an equation describing vesicle evolution in a fluid where some
stationary flow is excited regarding that the vesicle shape is close to a sphere. A
character of the evolution is governed by two dimensionless parameters, $S$ and
$\Lambda$, depending on the vesicle excess area, viscosity contrast, membrane viscosity,
strength of the flow, bending module, and ratio of the elongation and rotation components
of the flow. We establish the ``phase diagram'' of the system on the $S-\Lambda$ plane:
we find curves corresponding to the tank-treading to tumbling transition (described by
the saddle-node bifurcation) and to the tank-treading to trembling transition (described
by the Hopf bifurcation).

 \end{abstract}


\maketitle

Vesicles are closed membranes which separate two regions occupied by possibly different
fluids. The vesicles are attracting significant attention not only due to their
resemblance with biological objects but also because of their importance in different
industries such as pharmaceutics where they are used for drug transportation. A natural
problem which arises in these applications is understanding of how a single vesicle
behaves in an external flow. This non-equilibrium problem has revealed a variety of new
physical effects and became a subject of intense experimental and theoretical studies.
Laboratory experiments \cite{HBVD97,SBAM98,ALV02,KS05,MVAV06} have shown that the
vesicles immersed in a shear flow exhibit at least two qualitatively different types of
behavior, either tank-treading or tumbling motion. In the tank-treading regime a vesicle
shape is stationary, it is ellipsoid oriented at an angle with respect to the shear flow.
In the tumbling regime the vesicle experiences periodic flipping in the shear plane. A
novel type of behavior: trembling, discovered in the work \cite{KS06} is an intermediate
regime between tank-treading and tumbling in which a vesicle trembles around the flow
direction.

Constructing a phase diagram for all these regimes depending on the external parameters
is a challenging and an extremely difficult task because the problem in consideration is
both strongly non-linear and non-equilibrium. As long as no analytic solution of this
problem exists theoretical studies were based either on numerical simulations or on some
approximations allowing analytical treatment. Numerical investigations of this problem
involved several different computational schemes, including boundary element method
\cite{KWSL96_SS01}, mesoscopic particle-based approximation
\cite{NT02,NG04,NG05_1,NG05_2,NG05_3}, and an advected field approach
\cite{BM02_BM03_BRSBM04_BKM05}. These approaches have shown qualitative agreement with
experiments however did not solve the problem of constructing the vesicle dynamics phase
diagram completely. Analytical studies of the problem can be divided in two major
classes. In the first one \cite{RBM04,NG04,NG06}, phenomenological models of a vesicle
dynamics based on the classical work of Keller and Skallak \cite{82KS} were proposed and
proved themselves to be rather efficient in explaining the experiments. In the second
series of works \cite{S99,O00,M06,VG07} the studies focused on quasi-spherical vesicles
whose shape can be parameterized by spherical harmonics expansion. A perturbation scheme
around the Lamb solution for spherical particle in external flow allows one to derive the
dynamic equations for the shape and the orientation of a vesicle and to investigate them
analytically. In this letter we propose a natural extension of the theory developed for
quasi-spherical vesicles by accounting higher-order expansion terms. We show that these
additions produce a qualitative change in the phase diagram and make it significantly
more complicated. The resulting diagram which is the main result of this paper contains
all three types of vesicle behavior which were observed experimentally. In our work we
analyze how the vesicle dynamics depends on different control parameters, such as
viscosity contrast, vesicle excess area, internal membrane viscosity, strength of the
flow, and ratio of the elongation and rotation components of the flow. We analyze also
the vesicle orientation in the tank treading regime.

Speaking about membranes we have in mind lipid bilayers. Physical properties of such
objects have been extensively studied both experimentally and theoretically (see e.g. the
books \cite{ML87,SC87,NP89} and the reviews \cite{BP84_PA91_PO92}). There are several
features of the membranes which are important for our analysis. First, we assume that the
membrane is in a fluid state (is a $2d$ liquid), which is typical of the lipid bilayers
under normal conditions. Second, we assume that the vesicle has an excess area, that
enables one to treat the vesicle membrane as incompressible. Third, we assume that the
membrane is impermeable to the surrounding liquids, the condition is usually well
satisfied in experiment. Finally, we take into account the membrane internal viscosity,
which could play an essential role, say, in a vicinity of the lipid-bilayer melting point
\cite{DPD00} (see also Refs. \cite{NG04,NG05_3}).

We assume that the vesicle size is much larger than the membrane thickness. Then, in the
main approximation, the vesicle membrane can be treated as infinitesimally thin, that is
as a $2d$ object immersed into a $3d$ fluid. By other words, in the limit the vesicle can
be considered as an interface separating two generally different fluids. The two
properties, the membrane incompressibility and impermeability, imply that both the
vesicle volume ${\cal V}$ and the membrane area ${\cal A}$ are conserved. The excess area
can be characterized by a dimensionless factor $\Delta$, which is defined as ${\cal A} =
(4\pi+\Delta)r_0^2$, where $r_0$ is a vesicle ``radius'' determined by its volume: ${\cal
V} = 4\pi r_0^3/3$. The excess area is non-negative, $\Delta\geq0$, and the minimal value
$\Delta =0$ corresponds to ideal spherical geometry.

The membrane free energy can be written as the following integral over the membrane
position \cite{70Can,73Hel,74Eva,75Hel}
 \begin{equation}
 {\cal F}=\int dA \left(
 \sigma+\frac{\kappa}{2}H^2\right),
 \label{free}
 \end{equation}
where $\sigma$ is the membrane surface tension, $H$ is its mean curvature, and $\kappa$
is Helfrich module. The last term in Eq. (\ref{free}) describes energy related to
membrane bending distortions. Note that the surface tension $\sigma$ is a quantity
adjusting to other membrane parameters (similarly to pressure in incompressible fluid) to
ensure a given value of the membrane area $A$.

The membrane moves together with surrounding fluid that is the velocity field $\bm v$ is
continuous on the membrane and $\bm v$ determines the membrane velocity as well as the
fluid velocity. We divide the flow near the vesicle into two parts: an external flow
which would occur in the fluid in the absence of the vesicle, and an induced flow which
is excited as a result of the vesicle reaction to the external flow. We assume that a
characteristic scale of the external flow is much larger than the vesicle size. Then the
external flow velocity $\bm V$ near the vesicle can be approximated by a linear profile.
Generally, the external flow has both the strain (elongation) and the rotational
contributions: $\partial_i V_k = s_{ik} + \epsilon_{ikj} \omega_j$, where $\hat s$ is the
(symmetric) strain matrix and $\bm\omega$ is the angular velocity vector. The strain can
be characterized by its strength $s$, defined as $s^2= \mathrm{tr}\ \hat s^2/2$. Note
that for a shear flow $s=\omega=\dot\gamma/2$, where $\dot\gamma$ is the shear rate.

We examine nearly spherical vesicles, that is the excess area parameter $\Delta$ is
regarded to be small. Then it is natural to describe the vesicle shape (membrane
position) as $r = r_0[1 + u(\theta,\varphi)]$ where $r,\theta,\varphi$ are spherical
coordinates in the reference system with origin chosen at the vesicle center. The
quantity $u$ is a dimensionless displacement characterizing deviations of the membrane
shape from a spherical one. For small $\Delta$ the value of $u$ can be estimated as
$u\sim \sqrt\Delta$. Therefore in our scheme $u\ll1$ and one can formulate a perturbation
theory in the parameter.

We assume that both interior fluid and exterior one are Newtonian and that Reynolds
number is vanishingly small, so the fluids can be described in terms of Stokes equation.
We assume also that an adiabaticity condition $\mathrm{max}\{s,\kappa/(\eta r_0^3)\}
\ll\eta/(\varrho r_0^2)$ is satisfied where $\eta$ is viscosity and $\varrho$ is mass
density of the exterior fluid. The same condition is assumed for the interior fluid. Then
the flow inside and outside the vesicle can be treated as instantaneously adjusted to the
vesicle motion. In this case it is possible to establish a closed dynamic equation for the
membrane displacement $u$. For the purpose one should find the velocity field $\bm v$
inside and outside the vesicle at a given displacement $u(\theta,\varphi)$ and then
equate $\partial_t u$ to the membrane normal velocity. To find the velocity field one
should solve the Stokes equation with boundary conditions dictated by the membrane
incompressibility and by the momentum balance that includes membrane forces determined by
the energy (\ref{free}) \cite{89ZH,89LM,S99} and membrane viscosity. To realize the
program for a nearly spherical vesicle one can use a generalization of the Lamb scheme
applicable to a spherical body, see Ref. \cite{32Lamb}. As a result, one finds the
dynamic equation for $u$ as a series in $u$.

In the main approximation in $u$ one obtains
 \begin{eqnarray}
 \hat a(\partial_t-\omega\partial_\varphi)u =
 10 s_{ij}\frac{r_ir_j}{r^2}
 -\dfrac{1}{\eta r_0^3}
 \dfrac{\delta {\cal F}^{(3)}}{\delta u} \,,
 \label{main}
 \end{eqnarray}
where $\hat a$ is some dimensionless operator, reflecting all viscous mechanisms, the
$Z$-axis of our reference frame is chosen to be directed opposite to the angular velocity
$\bm\omega$, and ${\cal F}^{(3)}$ is an expansion up to third order in $u$ of the free
energy (\ref{free}). Note that the elongation and the rotation parts of the external flow
are separated in Eq. (\ref{main}): the angular velocity $\bm\omega$ extends the time
derivative whereas the strain determines the non-uniform term playing a role parallel to
the free energy derivative. For an external shear flow $s_{ij}r_ir_j/r^2 =
({\dot\gamma}/{2}) \sin^2\theta\,\sin(2\varphi)$, where $X$-axis is directed along the
velocity.

The two differences between the equation (\ref{main}) and the analogous equations
obtained by Misbah \cite{M06} and Vlahovska et. al. \cite{VG07} are the inclusion of
membrane viscosity and third order expansion term of Helfrich energy. An account of the
third order term changes the phase diagram completely since the second order term becomes
vanishingly small (due to the surface tension adjustment) in the vicinity of the tumbling
to tank-treading transition region predicted in Refs. \cite{M06,VG07}. An importance of
the third-order term is discussed also by Noguchi and Gompper \cite{NG06}.

We take into account only the main contribution to $u$ determined by second order angular
harmonics. It is justified by the smallness of $\Delta$ and by the fact that the
non-uniform term in the right-hand side of Eq. (\ref{main}) is a linear combination of
second order angular harmonics. The operator $\hat a$ in the case is reduced to a
constant
 \begin{equation}
 a=\frac{16}{3}
 \left(1+\frac{23}{32}\frac{\tilde\eta}{\eta}
 +\frac{\zeta}{\eta r_0}\right) \,,
 \label{defa}
 \end{equation}
where $\tilde\eta$ is viscosity of the interior fluid and $\zeta$ is membrane viscosity.
Note that the expression (\ref{defa}) includes the viscosity contrast $\tilde\eta/\eta$.
After passing to the variable $u/\sqrt\Delta$ the equation (\ref{main}) acquires a
self-similar form containing the parameters $\sqrt\Delta\, a \omega/s$, $s\eta
r_0^3/(\kappa\Delta)$ and some dimensionless parameters characterizing ratios of the
eigenvalues of the matrix $\hat s$ and mutual orientation of the vorticity vector $\bm
\omega$ and the main axes of the matrix $\hat s$.

Below, we consider a plain flow where the external velocity $\bm V$ lies in $X-Y$ plane.
In the case the equation for $u/\sqrt\Delta$ depends on two dimensionless parameters
which can be chosen as
 \begin{equation}
    \Lambda    =
    {\frac{\sqrt{3}}{4\sqrt{10\pi}}}\frac{\sqrt\Delta\, a \omega}{s},
    \quad
    S = \frac{14\pi}{3\sqrt{3}}\frac{s \eta r_0^3}{\kappa\Delta}.
 \label{express}
 \end{equation}
In the case of weak strains, $S\ll 1$, the vesicle conserves its equilibrium shape, and
in the case of strong strains, $S\gg1$, the vesicle shape is determined by the external
flow. The parameter $\Lambda$ determines an effectiveness of the rotational part of the
external flow. Note that our theory is applicable, particularly, to purely elongation
flow, where $\omega=0$ and, consequently, $\Lambda=0$.

For the plain flow, one can further reduce description of the vesicle shape which is
dependent on two parameters, $\Theta$ and $\Phi$:
 \begin{equation*}
 u\! =\frac{\sqrt{5\Delta}}{4\sqrt{2\pi}}
 \left[\frac{\sin\!\Theta}{\sqrt 3}(1\!-\!3\cos^2\!\theta)\! +\!
 \cos\!\Theta \sin^2\!\theta \cos\! 2(\varphi\!-\!\Phi)\right].
 \end{equation*}
Here, the axes $X$ and $Y$ are chosen to be directed along the principal axes of the
strain matrix $\hat s$. Note that the ``angles'' $\Phi,\Theta$ determine the main axis
direction of the vesicle projection to the $X-Y$ axis. In the limit $\Delta \ll 1$ the
inclination angle of the vesicle projection main axis is $\varphi^* =\Phi$ for
$\cos\Theta >0$ and $\varphi^* = \Phi+\pi/2$ otherwise. The parameter $\Theta$ determines
also the vesicle shape and can be directly related to the parameter $R = \cos\Theta$
introduced in Ref. \cite{M06}. In terms of the variables $\Phi$ and $\Theta$, Eq.
(\ref{main}) is reduced to a couple of equations
\begin{eqnarray}
 \label{general1}
    \tau\partial_t\Phi =
    \frac{S}{2}\left[\frac{\cos(2\Phi)}{\cos\Theta}
    -\Lambda\right],
  \\ \label{general2}
    \tau\partial_t\Theta=
    -S\sin\Theta\sin(2\Phi)+\cos(3\Theta),
 \\ \label{taustar}
 \mathrm{where} \qquad
 \tau = \frac{7\sqrt{\pi}}{12\sqrt{10}}
 \frac{a\eta r_0^3}{\kappa\sqrt\Delta}.
 \label{tau} \end{eqnarray}

 \begin{figure}
 \includegraphics[width=3.5in,angle=0]{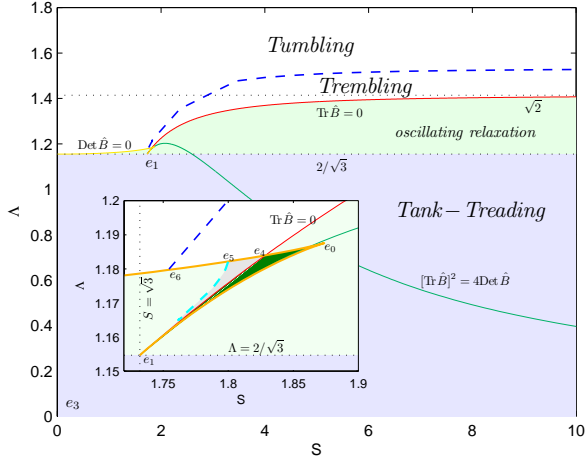}
 \caption{Phase diagram on the $S-\Lambda$ plane.}
 \label{fig:phase} \end{figure}

The tank treading vesicle motion corresponds to stable stationary points of the equations
(\ref{general1},\ref{general2}). Equating to zero the right hand sides of the equations
one finds relations determining a stationary point at given parameters $S$ and $\Lambda$.
To investigate stability of the point one should linearize the equations (\ref{general1},
\ref{general2}) near the point to obtain $\tau\partial_t(\delta\Theta,\delta\Phi)=\hat
B(\delta\Theta,\delta\Phi)$. The point is stable, if both eigenvalues of the matrix $\hat
B$ have negative real parts. Thus the stability conditions are $\mathrm{tr}\ B<0$ and
$\mathrm{det}\ B>0$. Regions in the $S-\Lambda$ plain where stable points exist are
indicated in Fig. \ref{fig:phase} by blue color. Note that there exists an additional
region of stationary points which are stable in terms of $\Theta$ and $\Phi$. However, a
stability investigation in the framework of Eq. (\ref{main}) shows that the points are
unstable in the extended space where the spherical harmonics $Y_{2,\pm 1}(\theta,\phi)$
are involved into play. Therefore we do not analyze the points.

If $\Lambda$ is increased at a given value of $S$ then first the stable point is in the
domain where $\Phi>0$. At the boundary of the domain, at $\Lambda=2/\sqrt3$, the
parameters $\Phi$ and $\Theta$ come to the values $\Phi=0$, $\Theta=\pi/6$ . Then the
stable point moves into the domain, corresponding to $\Lambda>2/\sqrt3$ and $\Phi<0$. The
green line in Fig. \ref{fig:phase} is determined by the condition $(\mathrm{tr}\,
B)^2=4\,\mathrm{det}\, B$, it separates regions where oscillatory and damping regimes of
approaching the stable point occur. The boundary of the stability domain can be
determined either by the condition $\mathrm{det}\, B=0$ or $\mathrm{tr}\, B=0$. The
corresponding lines are depicted in Fig. \ref{fig:phase} by yellow and red lines,
respectively. The first case is realized at small $S$ whereas the second case is realized
at large $S$. At increasing $\Lambda$ a Hopf bifurcation occurs at the red line whereas a
saddle-node bifurcation occurs at the yellow line. A limit cycle is realized in the
system for points above the red and the yellow lines (that is at large $\Lambda$).
Following the work \cite{KS06} we distinguish tumbling and trembling regimes. In our
terms, trembling corresponds to a limit cycle where $\varphi$ varies in a restricted
interval whereas tumbling means an increase of $\varphi$ by $2\pi$ during the cycle. The
dashed blue line in Fig. \ref{fig:phase} separates the two regimes, trembling is realized
below the line. A possibility of such regime is discussed also by Misbah \cite{M06}. The
sequence of the vesicle states predicted by our theory corresponds to one observed in
experiment \cite{KS06}. Qualitative the picture is also similar to the one observed in
recent numerical simulations by Noguchi and Gompper \cite{NG06}. 
Note that at infinitely large $\tilde\eta$ or $\zeta$ a solid
ball behavior of the vesicle should be observed which is tumbling as Jeffery established
\cite{jeffery}). In our scheme large $\tilde\eta$ or $\zeta$ implies large $\Lambda$,
corresponding to tumbling region, indeed.

A special point $S=\sqrt3$, $\Lambda=2/\sqrt3$, where two transition lines terminate, is
analogous to some extend to fluid critical point. Say, frequency of the Hopf bifurcation
behaves $\propto\sqrt{S^2-3}$. Similar ``smoothing'' of the saddle-node bifurcation near
the point occurs. A complicated phase structure is observed near the special point, its
narrow vicinity is plotted in the insertion to Fig. \ref{fig:phase}. The green region
corresponds to coexistence of two stable points. There exists also a region  where the
stable point and the limit cycle coexist. The limit cycle becomes unstable at the dashed
green line.

Thermal fluctuations of the vesicle shape can be investigated in spirit of Refs.
\cite{96KLM,S99,07Tur} (see also the book \cite{93KL}). The fluctuations produce
isotropic Brownian motion on the sphere, defined by the ``angles'' $\Theta$ and $\Phi$
with the characteristic angular diffusion coefficient $D\sim k T /(\eta r_0^3 \Delta)$,
where ${T}$ is temperature. In the tank-treading regime and strong external flow,
$S\gg1$, fluctuations induced by this diffusion can be estimated as $(\delta
\Theta)^2\sim (\delta \Phi)^2 \sim D \tau/S$ if $\Lambda<2/\sqrt3$ and as $(\delta
\Theta)^2\sim (\delta \Phi)^2 \sim D \tau$ if $\Lambda>2/\sqrt3$, where $\tau$ is defined
by Eq. (\ref{tau}). For weak external flow, $S\ll1$, fluctuations of vesicle shape are
given by estimate $(\delta \Theta)^2\sim D \tau$, whereas vesicle orientation experiences
more stronger fluctuations, $(\delta \Phi)^2 \sim D \tau/S$. The thermal fluctuations
play an essential role near the tank-treading to tumbling (or tank-treading to trembling)
transition ``smearing'' it. Results of detailed investigation of this phenomenon will be
published elsewhere.

To conclude, we have theoretically investigated the vesicle dynamics in an external
stationary flow. The general scheme based on solving the $3d$ hydrodynamic (Stokes)
equations with boundary conditions posed on the membrane enabled us to analyze in detail
dynamical properties of nearly spherical vesicles. We constructed the phase diagram of
the system in terms of two dimensionless parameters (\ref{express}) which are
combinations of physically observed quantities. We demonstrated that there are two
different regimes realized in weak and strong external flows. In weak flows the vesicle
shape is close to an equilibrium one which is a prolate ellipsoid and a role of the
external flow is reduced mainly to an orientation of the ellipsoid. Then the
tank-treading to tumbling transition occurs which can be described by a saddle-node
bifurcation. In strong flows the vesicle shape and orientation are determined by the
flow. Then the tank-treading to trembling transition occurs which can be described by a
Hopf bifurcation. The two transition curves are separated by a point vicinity of which
needs a special consideration.

The authors acknowledge numerous discussions of the experiments with V. Steinberg and V. Kantsler.
This work has been partially supported by RFBR. KT and SV also acknowledge the financial
support  from ``Dynasty'' and RSSF foundation.


\begin{thebibliography}{99}

 \bibitem{HBVD97}
 K. H. De Haas, C. Blom, D. E. Van, D., M. H. G. Duits, J. Mellema,
 Phys. Rev. E \textbf{56}, 7132 (1997).

 \bibitem{SBAM98}
 N. Shahidzadeh, D. Bonn, O. Aguerre-Chariol, J. Meunier,
 Phys. Rev. Lett. \textbf{81}, 4268 (1998).

 \bibitem{ALV02}
 M. Abkarian, C. Lartigue, and A. Viallat,
 Phys. Rev. Lett. \textbf{88}, (2002).


 \bibitem{KS05}
 V. Kantsler and V. Steinberg,
 Phys. Rev. Lett., \textbf{95}, 258101 (2005).


 \bibitem{MVAV06}
 M. A. Mader, V. Vitkova, M. Abkarian, A. Viallat, and T. Podgorski,
 Eur. Phys. J. E \textbf{19}, 389 (2006).

 \bibitem{KS06}
 V. Kantsler and V. Steinberg,
 Phys. Rev. Lett., \textbf{96}, 036001 (2006).

 \bibitem{KWSL96_SS01}
 M. Kraus, W. Wintz, U. Seifert, and R. Lipowsky,
 Phys. Rev. Lett. \textbf{77}, 3685 (1996).
 S. Sukumaran and U. Seifert, Phys. Rev. E, \textbf{64}, 011916 (2001).

 \bibitem{NT02}
 H. Noguchi and M. Takasu, Phys. Rev. E \textbf{65}, (2002).

 \bibitem{NG04}
 H. Noguchi and G. Gompper,
 Phys. Rev. Lett., \textbf{93}, 258102 (2004)

 \bibitem{NG05_1}
 H. Noguchi and G. Gompper, Journal of Physics Condensed Matter,
 \textbf{17}, S3439 (2005)

 \bibitem{NG05_2}
 H. Noguchi and G. Gompper,
 Proc. Nat. Ac. Sci., \textbf{102}, 14159-14164 (2005)

 \bibitem{NG05_3}
 H. Noguchi and G. Gompper, Phys. Rev. E ., \textbf{72}, 011901 (2005)

 \bibitem{BM02_BM03_BRSBM04_BKM05}T. Biben, C. Misbah, Eur. Phys. J B \textbf{29}, 311 (2002).
 T. Biben, C. Misbah, Phys. Rev. E, \textbf{67}, 031908 (2003). 
 J. Beaucourt, F. Rioual, T. Seon, T. Biben, and C. Misbah,
 Phys. Rev. E {\bf 69}, 011906 (2004). 
 T. Biben, K. Kassner, and C. Misbah,
 Phys. Rev. E \textbf{72}, (2005). %

 \bibitem{RBM04}
 F. Rioual, T. Biben, and C. Misbah, Phys. Rev. E \textbf{69}, (2004).

 \bibitem{NG06}
 H. Noguchi and G. Gompper,
 \textit{Swinging and Tumbling of Fluid Vesicles in Shear Flow},
 arXiv:cond-mat/0611382

 \bibitem{82KS}
 S. R. Keller and R. Skalak,
 J. Fluid Mech. {\bf 120}, 27 (1982).

 \bibitem{S99}
 U. Seifert, Eur. Phys. J. B, \textbf{8}, 405 (1999).

 \bibitem{O00}
 P. Olla, Physica A,
 \textbf{278}, 87-106 (2000).

 \bibitem{M06}
 C. Misbah, Phys. Rev. Lett., \textbf{96}, 028104 (2006).

 \bibitem{VG07}
 P. M. Vlahovska and R. S. Gracia, Phys. Rev. E \textbf{75}, (2007).

 \bibitem{ML87}
 Physics of Amphiphilic Layers,
 J. Meuner, D. Langevin, and N. Boccara,
 Springer Proceedings in Physics, {\bf 21},
 Springer-Verlag, Berlin, 1987.

 \bibitem{SC87}
 S. A. Safran and N. A. Clark,
 Physics of Complex and Supermolecular Fluids,
 Wiley, NY, 1987.

 \bibitem{NP89}
 D. Nelson, T. Pvian, and S. Weinberg,
 Statistical Mechanics of Membranes and Surfaces,
 World Scientific, NY, 1989.

 \bibitem{BP84_PA91_PO92}
 A. M. Bellocq et. al., Adv. Colloid Interface Sci. {\bf 20}, 167 (1984). 
 G. Porte, et. al., Physica A{\bf 176}, 168 (1991).
 G. Porte, et. al. J. Phys. II  {\bf 4}, 8649 (1992).

 \bibitem{DPD00}
 R. Dimova, B. Pouligny, and C. Dietrich,
 Biophys. J. {\bf 79}, 340 (2000).

 \bibitem{70Can}
 P. B. Canham, J. Theor. Biol. {\bf 26}, 61 (1970).

 \bibitem{73Hel}
 W. Helfrich, Z. Naturforsch. A {\bf 28c}, 693 (1973).

  \bibitem{74Eva}
 E. Evans, Biophys. J. {\bf 14}, 923 (1974).

  \bibitem{75Hel}
 W. Helfrich, Z. Naturforsch {\bf B103}, 67 (1975).

 \bibitem{89ZH}
 Ou-Yang Zong-Can and W. Helfrich,
 Phys. Rev. A{\bf 39}, 5280 (1989).

  \bibitem{89LM}
 V. V. Lebedev and A. R. Muratov,
 ZhETF {\bf 95}, 1751 (1989)
 [Sov. Phys. JETP {\bf 68} 1011 (1989)].

  \bibitem{32Lamb}
 H. Lamb, Hydrodynamics (Cambridge Uiversity Press, Cambridge, England, 1932), 6th ed.

  \bibitem{jeffery}
 G. B. Jeffery,
 Proceedings of the Royal Society of London. Series A,
 {\bf 102}, No. 715, 161-179 (1922).

  \bibitem{96KLM}
 E. I. Kats, V. V. Lebedev, and A. R. Muratov,
 Nearly spherical vesicles: Shape fluctuations,
 Pis'ma v ZhETF, {\bf 63}, 203 (1996)
 [JETP Lett. {\bf 63}, 216-221 (1996)].

  \bibitem{07Tur}
 K. S. Turitsyn, submitted to ZhETF, arXiv:nlin.CD/0501025.

  \bibitem{93KL}
 E. I. Kats and V. V. Lebedev,
 Fluctuational Effects in the Dynamics of Liquid Crystals
 Springer-Verlag, NY, 1993.

\end{thebibliography}
\end{document}